\documentclass[twocolumn,pre]{revtex4}
\newcommand{\figurewidth}{84mm}

\usepackage{graphicx}
\usepackage{dcolumn}
\usepackage{amsmath}

\newcommand{\leqs}{\stackrel{\scriptstyle<}{\scriptscriptstyle\sim}\:}

\newcommand{\PP}{{\mathcal{P}}}
\newcommand{\NN}{{\mathcal{N}}}

\newcommand{\RR}{{\bf R}}
\newcommand{\RRp}{{\bf R'}}

\newcommand{\rr}{{\bf r}}

\newcommand{\mm}{{\bf m}}

\newcommand{\HH}{{\mathcal{H}}}

\newcommand{\BE}{\begin{equation}}
\newcommand{\EE}{\end{equation}}
\newcommand{\BEN}{\begin{eqnarray}}
\newcommand{\EEN}{\end{eqnarray}}

\begin{document}


\title[Short Title]{
                    Path Integral Monte Carlo Simulation of the Low-Density Hydrogen Plasma}

\author{B. Militzer}
\email{militzer@llnl.gov}
\affiliation{Lawrence Livermore National Laboratory,
         University of California, Livermore, CA 94550}

\author{D. M. Ceperley}
\email{ceperley@uiuc.edu} \affiliation{Department of Physics,
             National Center for Supercomputing Applications,
             University of Illinois at Urbana-Champaign, Urbana, IL 61801}

\date{\today }

\begin{abstract}
Restricted path integral Monte Carlo simulations are used to
calculate the equilibrium properties of hydrogen in the density
and temperature range of $9.83 \times 10^{-4}\rm \leq \rho \leq
0.153\,\rm gcm^{-3}$ and $5000 \leq T \leq 250\,000\,\rm K$. We
test the accuracy of the pair density matrix and analyze the
dependence on the system size, on the time step of the path
integral and on the type of nodal surface. We calculate the
equation of state and compare with other models for hydrogen valid
in this regime.  Further, we characterize the state of hydrogen
and describe the changes from a plasma to an atomic and molecular
liquid by analyzing the pair correlation functions and estimating
the number of atoms and molecules present.
\end{abstract}


\maketitle


\section{Introduction}

\begin{figure}[htb]
\includegraphics[angle=0,width=\figurewidth]{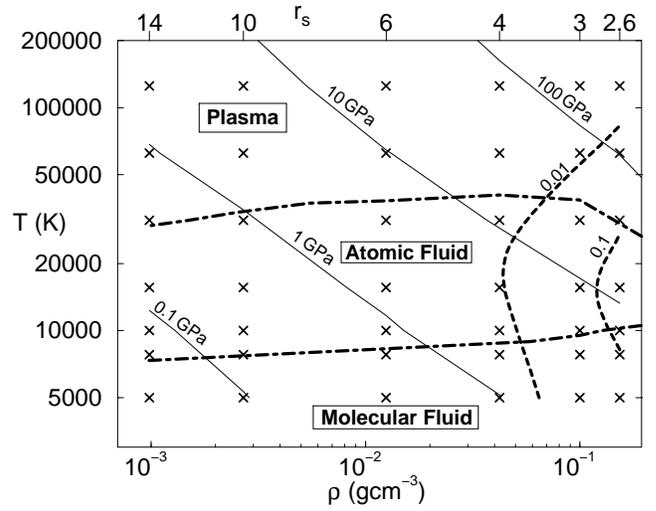}
\caption{Phase diagram of hydrogen as a function of temperature
         and density for the different regimes: plasma, the atomic and
         the molecular regime. The dash-dot lines indicate the
         approximate boundaries. The crosses indicate the parameters
         for which PIMC simulations have been performed, the solid
         lines are isobars, and the dashed lines represent contour
         lines of constant permutation probability of the electrons
         (as indicated on the line.) }
\label{phase}
\end{figure}

In spite of the simple composition, hydrogen exhibits a
surprisingly complex phase diagram, which is the subject of
numerous experimental and theoretical approaches. In this work, we
study the high temperature regime of $5000 \leq T \leq
250\,000\,\rm K$ where hydrogen undergoes a smooth transition with
increasing temperature from a molecular fluid through an atomic
regime and finally to a two component plasma of electrons and
protons (see Fig.~\ref{phase}).
The properties of hydrogen in this regime are crucial for the
evolution of stars and the characteristics of the Jovian planets.

A variety of simulation techniques and analytical models have been
developed to describe hydrogen at low density.  This regime has
been studied with chemical models \cite{SC92,ER85b,Ki97} that describe
hydrogen as a mixture of interacting molecules, atoms, free
protons and electrons. The chemical composition is determined by
minimizing an approximate free energy function constructed from
known theoretical limits. In this paper, we focus on low and
intermediate densities
$9.83 \times 10^{-4}\rm \leq \rho \leq
0.153\,\rm gcm^{-3}$ corresponding to $14 \geq r_s \geq 2.6$,
where one expects the chemical models to work well although the
properties of hydrogen are determined by the complex interplay of
long-range Coulomb forces leading to strong coupling and bound states
as well as degeneracy effects.

All these effects can also be described from first principles
simulation. There are {\em ab initio} methods such as restricted
path integral Monte Carlo simulations (PIMC) \cite{PC94,Ma96,Mi99}
and density functional theory molecular dynamics (DFT-MD)
\cite{Le00,Ga99}. The focus of the work is to test the equation of
state (EOS) derived from chemical models and the actual
density-temperature limits of the validity of the chemical
picture. Additionally, we provide data to determine the parameters
of the free energy models. Chemical models are expected to become
inaccurate in regions of high density where the lifetime of
molecules reduces to a few molecular vibrations \cite{Ga99}.

We present results from new, more accurate, PIMC simulations.
First, we analyze the accuracy of the pair density matrix and the
error of the ``time'' discretization. Then we analyze the finite
size dependence of the derived EOS and discuss the fixed node
errors by comparing results from simulations using free particle
(FP) nodes and variational nodes. Furthermore, we calculate pair
correlation functions which we use in conjunction with a cluster
analysis to characterize the state of hydrogen at different
temperatures and densities.  We use atomic units (lengths in Bohr
radii and energies in Hartrees) throughout this work except where
indicated otherwise.

\section{Path Integral Monte Carlo Method}

\subsection{Restricted Path Integral}

The density matrix (DM) of a fermion system at temperature $k_BT=1
/ \beta$ can be written as an integral over all paths $\RR_t$,

\begin{equation}
\rho(\RR_0 ,\RR_\beta; \beta) =
   \frac{1}{N!} \sum_\PP \:
   (- 1)^\PP \! \! \! \! \! \oint\limits_{\RR_0 \rightarrow \PP\RR_{\beta}}
    \! \! \! \! \! \! d\RR_t \;\; e^{-S[\RR_t]}.
\end{equation}

$\RR_t$ stands for the entire paths of $N$ particles in $3$
dimensional space $\RR_t = (\rr_{1t},\ldots,\rr_{Nt})$ beginning
at $\RR_0$ and ending at $\PP \RR_\beta$. $\PP$ labels the
permutation of the particles and $(- 1)^\PP $ to its signature.
For non-relativistic particles interacting with a potential
$V(\RR)$, the action of the path $S[\RR_t]$ is given by,

\begin{equation}
S[\RR_t] =
\int_0^\beta \! dt \left[ \frac{m}{2} \left| \frac{d\RR(t)}{\hbar
dt} \right|^2+ V(\RR(t)) \right] + \mbox{const}.
\label{primitive_action}
\end{equation}
In practice one discretizes \cite{Ce95} the path into a finite
number of imaginary time slices $M$ corresponding to a {\em time
step} $\tau=\beta/M$.

For fermionic systems the integration is complicated due to the
cancellation of positive and negative contributions to the
integral, ({\em the fermion sign problem}). It has been
shown~\cite{Ce91,Ce96} that one can evaluate the path integral by
restricting the path to only specific positive contributions. One
introduces a reference point $\RR^*$ on the path that specifies
the nodes of the DM, $\rho(\RR,\RR^*,t)=0$. A {\em node-avoiding}
path for $ 0 < t \leq \beta $ neither touches nor crosses a node:
$\rho(\RR(t),\RR^*,t) \not= 0$.  By restricting the integral to
node-avoiding paths,
\begin{eqnarray}
\label{RPI} \nonumber
\rho_F(\RR_{\beta} ,\RR^*;\beta) &=&\\
\int \!\! d\RR_0 \: &\rho_F&(\RR_0, \RR^* ; 0)
\! \! \! \! \! \! \! \! \! \! \!
\oint\limits_{\RR_0 \rightarrow \RR_{\beta}\in \Upsilon(\RR^*)}
\! \! \! \! \! \! \! \! \! \! \! \! \!
d\RR_t \;\; e^{-S[\RR_t] },
\end{eqnarray}
($\Upsilon (\RR^*)$ denotes the restriction) the contributions are
positive and therefore PIMC represents, in principle, a solution
to the sign problem. The method is exact if the exact fermionic DM
is used for the restriction.  However, the exact DM is only known
in a few cases. Most applications have approximated the fermionic
DM by a determinant of single particle DMs,

\begin{equation}
\rho(\RR,\RRp;\beta)=\left|
\begin{array}{ccc}
\rho_1(\rr_{1},\rr'_{1};\beta)&\ldots&\rho_1(\rr_{N},\rr'_{1};\beta)\\
\ldots&\ldots&\ldots\\
\rho_1(\rr_{1},\rr'_{N};\beta)&\ldots&\rho_1(\rr_{N},\rr'_{N};\beta)
\end{array}\right|
\quad.
\label{matrixansatz}
\end{equation}

This approach has been extensively applied using the free particle
(FP) nodes derived from the single-particle density
matrix~\cite{Ce96}:
\begin{equation}
 \label{freegauss}
 \rho_1(\rr,\rr',\beta) = (4 \pi \lambda \beta)^{-3/2} \: \mbox{exp}
 \left\{ -(\rr-\rr')^2/4 \lambda \beta \right\}
\end{equation}
with $\lambda = \hbar^2/2m$, including applications to dense
hydrogen \cite{PC94,Ma96,Mi99}. It can be shown that for
temperatures larger than the Fermi energy, the interacting nodal
surface approaches the FP nodal surface. In addition, in the limit
of low density, exchange effects are negligible: the nodal
constraint has a small effect on the path and therefore its
precise shape is not important. The FP nodes also become exact in
the limit of high density when kinetic effects dominate over the
interaction potential.  However, for the densities and
temperatures under consideration, interactions could have a
significant effect on the nodal surfaces.

To gain some quantitative estimate of the possible effect of the nodal
restriction on the thermodynamic properties, it is necessary to try an
alternative. In addition to FP nodes, we used nodal surface of a
variational density matrix (VDM) \cite{MP00} derived from a variational
principle that includes interactions and atomic and molecular bound
states.  We assume a trial DM with parameters $q_i$ that depend on
imaginary time $\beta$ and $\RRp$,
\begin{equation}
\rho(\RR,\RRp;\beta)=\rho(\RR,q_1,\ldots,q_m)
\;.
\end{equation}
By minimizing the integral:
\begin{equation}
\int \! d\RR \left(
\frac{\partial \rho(\RR,\RRp;\beta)}{\partial \beta}+\HH \,
\rho(\RR,\RRp;\beta) \right)^{\!\!2} =0\quad, \label{varprinc}
\end{equation}
one determines equations for the dynamics of the parameters in
imaginary time:
\begin{equation}
   \label{matrix}
    \frac{1}{2}\frac{\partial H}{\partial \vec{q}}\; + \;\,
    \stackrel{{\textstyle \leftrightarrow}}{\NN}\: \dot{\vec{q}}=  0
    \;\;\;\;\mbox{where}\;\;\;\;
    H \equiv \int \rho \HH \rho\;d\RR  \;.
\end{equation}
The norm matrix is:
\begin{eqnarray}
\NN_{ij} &=&
\lim_{q'\rightarrow q} \frac{\partial^{\,2}}{ \partial q_i
\partial q'_j}\left[\int \!d\RR \;
\rho(\RR,\vec{q}\,;\beta) \; \rho(\RR,\vec{q}\:'\,;\beta)\right]
\;.
\label{NN}
\end{eqnarray}
We assume the DM is a Slater determinant of single
particle Gaussian functions
\begin{equation}
 \label{gauss}
 \rho_1(\rr,\rr',\beta) = (\pi w)^{-3/2} \: \mbox{exp}
 \left\{ -(\rr-\mm)^2/w + d \right\}
\end{equation}
where the variational parameters are the mean $\mm$, squared width $w$
and amplitude $d$. The differential equations for this ansatz are
given in~\cite{MP00}. The initial conditions at $\beta\longrightarrow
0$ are $w= 2\beta$, $\mm=\rr'$ and $d=0$ in order to regain the
correct FP limit.  It follows from Eq.~\ref{varprinc} that at low
temperature, the VDM goes to the lowest energy wave function within
the variational basis. For an isolated atom or molecule this will be a
bound state, in contrast to the delocalized state of the FP DM. A
further discussion of the VDM properties is given in \cite{MP00}.
Note that these nodes are only used to determine the nodal restriction
in Eq. (II.3). The complete potential is taken into account in the
path integral action as discussed in detail in \cite{Ce95}.

\subsection{Accuracy of the Method}
\label{Accuracy_of_the_Method}

The numerical implementation of the PIMC method requires one to
make several approximations. Inaccuracies can be caused by
statistical errors from the MC integration, inaccuracies in the
numerically determined pair density matrices, a dependence on the
time step of the path integral because of $N$-body ($N \geq 3$)
correlations, finite size effects, and nodal errors from
approximations in the trial density matrix. Their effects on the
accuracy of the computed thermodynamic averages are quantitatively
estimated in this section.

Statistical errors in the estimators for the thermodynamic
quantities are calculated from the block averages generated by the
MC simulations. The correlations between blocks are taken into
account by performing a blocking analysis \cite{AT87}. The
resulting error bars (one standard deviation) are given for all
observables throughout this paper as a number in parenthesis
referring to the least significant digit.

In the following discussion, we compare the internal energy and
the pressure calculated using the virial theorem:
\begin{equation}
\label{virial} 3 P v = 2K+V
\end{equation}
where $v$ is the volume of the simulation cell, $K$ the kinetic
energy and $V$ the potential energy. Accurate estimation of the
pressure requires a high accuracy in the kinetic and the potential
energies because they tend to cancel out. In the molecular regime
at low density, both terms, dominated by intra-molecular
contributions, cancel to a large extent leaving behind the
molecular gas pressure, which is much less than the
inter-molecular forces. As a result, the pressure is, in general,
significantly more sensitive to approximations than other
quantities such as the internal energy.

\subsubsection{Pair Density Matrix}

\begin{figure}[htb]
\includegraphics[angle=0,width=\figurewidth]{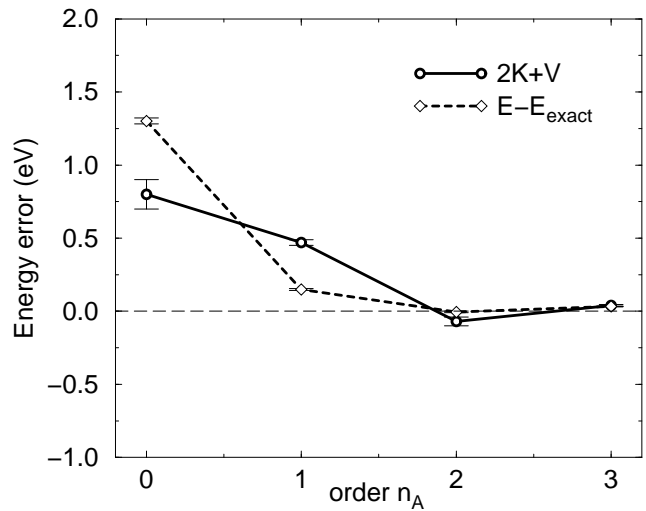}
\caption{Accuracy study of PIMC simulations of the isolated
     hydrogen atom using different orders $n_{\rm A}$ in the expansion
     formula~\ref{u_exp} for the action and energy. The
     calculated $2K+V$ (exact value equals zero) and the deviation
     from the exact potential energy of $-27.2\,\rm eV$ are shown for
     different orders from simulations at $T=10\,000\,\rm K$ using
     $\tau^{-1}=10^6\rm K$.}
\label{pair_dm_test_atom}
\end{figure}

If one only used the bare potential as in Eq.~\ref{primitive_action} (the primitive
approximation for the action), the convergence would be very
slow~\cite{BM00} and would result in an extremely inefficient
many-particle simulations. Instead, we numerically solve the
two-particle problem with the matrix squaring
technique~\cite{St68}. Numerical representations of the exact pair
density matrices are stored in tables used by the PIMC simulation
program by expanding in the small variables $s$ and $z$:
\begin{eqnarray}
\nonumber
u(\rr,\rr';\tau) = \frac{1}{2}
&&
\left[ u_0(r;\tau) + u_0(r';\tau) \right]
\\&&
+ \sum_{k=1}^{n_{\rm A}} \sum_{j=0}^{k} u_{kj}(q;\tau) \, z^{2j} \, s^{2(k-j)}
,
\label{u_exp}
\end{eqnarray}
where
\begin{equation}
q = \frac{1}{2}( |\rr|+|\rr'|)
\quad
s = |\rr-\rr'|
\quad
z = |\rr| - |\rr'|
\quad,
\end{equation}
and $\rr$ and $\rr'$ denote the separation of the two particles at
adjacent time slices. The accuracy of these tables is crucial for all
computed results. Using the precomputed pair density matrices allows
one to employ a much larger time step because one starts with a
solution of the two-particle problem.  Fig.~\ref{pair_dm_test_atom}
shows how accurate this method is.  The internal energy of an isolated
hydrogen atom at sufficiently low temperature ($T=10\,000\,\rm K$) in
a large box ($L=26$) is compared with the exact ground state energy of
$-13.6\,\rm eV$.  The temperature was chosen low enough so that
excited states can be neglected; the contribution to the energy from
the occupation of the first excited state is $7 \cdot 10^{-5}\,\rm eV$
at this temperature. Also shown is how well the kinetic energy $K$ and
the potential energy $V$ satisfy the virial theorem $2K+V=0$, thus
determining the accuracy with which the pressure can be determined.
For a time step of $\tau^{-1}=10^6\rm K$, the analysis shows a
quick convergence with the order of terms considered in the action
expansion Eq.~(\ref{u_exp}). Using terms up $n_{\rm A}=3$ reduces the
error to $0.033\,(3)\,\rm eV$ in energy and to $0.039\,(8)\, \rm eV$
for $2K+V$. This corresponds to an inaccuracy in the pressure
equivalent to a non-interacting molecular gas at $T=260\,(30)\,\rm K$.

\subsubsection{Time step dependence}

\begin{figure}[htb]
\includegraphics[angle=0,width=\figurewidth]{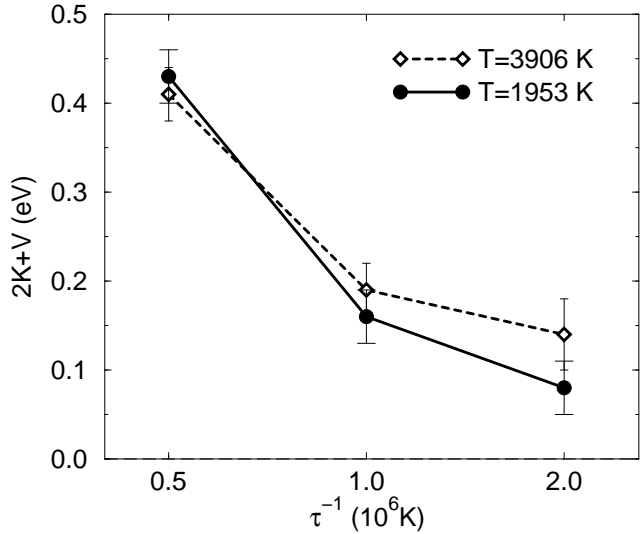}
\caption{The error in the virial as a function of the number of
         time slices for an isolated hydrogen molecule. The classical
         nuclei are fixed at the equilibrium bond length, thus the
         exact value for $2K+V$ is zero at a sufficiently low
         temperature.}
\label{H2_timestep}
\end{figure}

Employing the pair approximation of the density matrix does not
include correlation effects for three or more particles (for
example between one electron and two protons). We now estimate how
small the time step must be to obtain a given accuracy for an
isolated molecule. Fig.~\ref{H2_timestep} shows results for
different time steps and temperatures with the nuclei kept fixed
at the equilibrium position of $R=1.4008$. From the virial
theorem, it follows that $2K+V=0$ at a sufficiently low
temperature. The exact energy per atom is $-15.973\rm\,eV$
\cite{KW64}. The $T$ dependence is small suggesting that the
electrons are in the ground state. However, one finds a
significant dependence on the time step. Using $\tau^{-1}=2 \times
10^6\rm K$ reduces the error in the energy per atom to $0.036 (3)
\,\rm\,eV$ and in $2K+V$ to $0.090\,(16)\,\rm eV$. The time step
error is larger than the errors of the inaccuracies in the pair
density matrices discussed above. The error in $2K+V$ corresponds
to a pressure of an non-interacting molecular gas at
$T=700\,(100)\,\rm K$, which provides us with an approximate limit
of accuracy in the equation of state calculated from many-particle
simulations with $\tau^{-1}=2 \times 10^6\rm K$.

\subsubsection{Finite Size Dependence}

The estimation of the finite size errors is more difficult to
assess because the needed PIMC simulations are computationally
much more demanding. The required computer time increases rapidly
with the number of particles making it challenging to obtain
converged results for paths corresponding to large systems.

Most results from many-body simulations reported in this work were
calculated with $N=32$ pairs of electrons and protons in a
periodically repeated simulation cell. To study the effect on $N$,
we performed simulations for $N=16$ and $64$ pairs of protons and
electrons for a density of $r_s=2.6$ and $T \geq 10\,000\,\rm K$.
We chose the highest density under consideration because one
expects the finite size dependence to be largest there due to the
stronger interaction between the atoms.

\begin{figure}[htb]
\includegraphics[angle=0,width=\figurewidth]{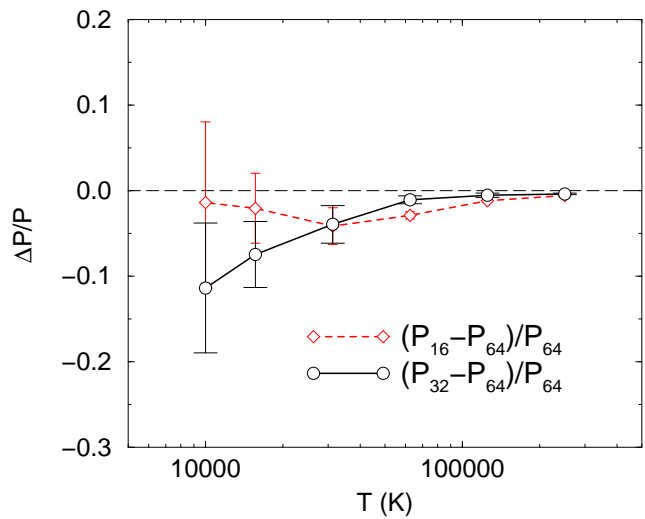}
\caption{Finite size error of the pressure as a function of temperature
     relative to simulations with $N=64$ pairs of protons and
     electrons at a density of $r_s=2.6$.}
\label{pressure_size}
\end{figure}

The finite size dependence of the pressure, shown in
Fig.~\ref{pressure_size}, is small at high temperatures but grows
to approximately $4\,(2)\,\%$ near $T=30\,000\,\rm K$. In this
regime, the hydrogen undergoes structural changes involving the
formation of atoms, which affect the pressure. This study provides
us only with an estimate of the finite size dependence. An
extrapolation to $N \to \infty$ would require significantly larger
systems, not currently feasible at low temperatures.

\begin{figure}[htb]
\includegraphics[angle=0,width=\figurewidth]{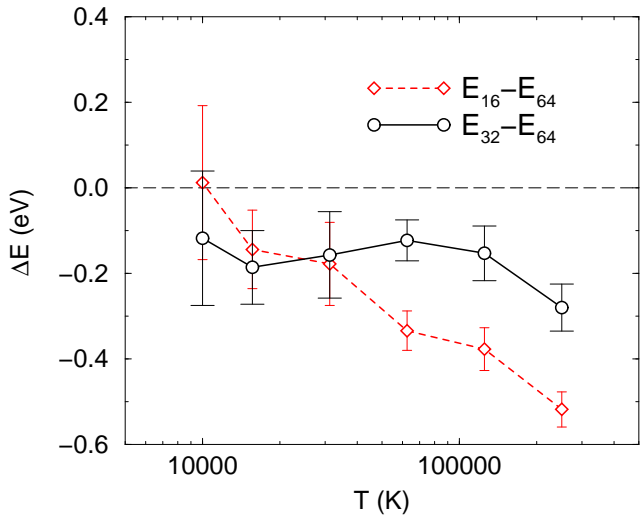}
\caption{Finite size error of the internal energy as a function of temperature
     relative to simulations with $N=64$ pairs of particles at a density of $r_s=2.6$.}
\label{energy_size}
\end{figure}

Fig.~\ref{energy_size} shows the finite size error of the internal
energy.
The smaller systems are more strongly bound by approximately
$0.2~\rm eV$ per atom, probably because of the interaction of a
charge with its own image. For lower densities, we expect this
value to be smaller.

\subsubsection{Nodal Approximation}

\begin{figure}[htb]
\includegraphics[angle=0,width=\figurewidth]{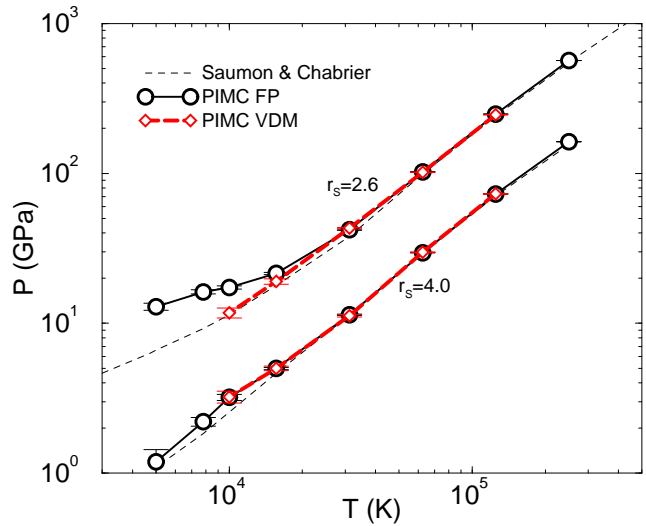}
\caption{The effect on the pressure of two
         different nodal surfaces: of the free particle density
         matrix and of the variational density matrix.}
\label{pressure_nodes}
\end{figure}

In the above, we have studied controlled approximations. The only
uncontrolled approximation in the restricted PIMC method is the
use of trial density matrix to constrain the paths. The nodal
surfaces are important only if the electrons are degenerate: at
low temperatures or at high densities. Recall that in this work we
focus on hydrogen only at low density, where the electrons are
bound in atoms and molecules and have a low or moderate
degeneracy. Even at low density, one still needs a nodal surface
in order to prevent the formation of unphysical clusters like
$H_3$ and $H_4$ or even the collapse of the entire system, but the
precise shape of the nodes is not important at low density as
shown in Fig.~\ref{pressure_nodes}.
Comparing FP and VDM nodes for $r_s=2.6$, one only finds
differences in the pressure for $T \leq 15\,625\,\rm K$, which is
approximately where, at this density, the system shows a
significant molecular signature (see
section~\ref{Pair_Correlation_Functions} and Fig.~\ref{grpp}).  In
this regime, FP nodes systematically lead to a too high pressure,
while simulations with VDM nodes stay closer to the prediction of
a semi-empirical chemical model \cite{SC92}.
At a lower density, $r_s=4$, results from FP and VDM nodes agree
within the error bars.
\begin{figure}[htb]
\includegraphics[angle=0,width=\figurewidth]{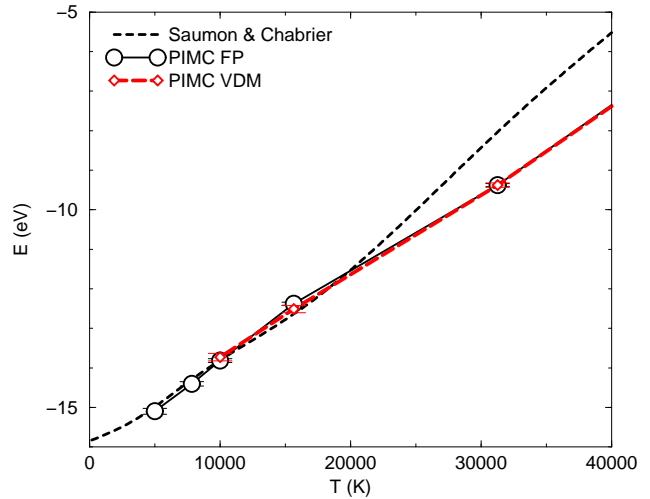}
\caption{Internal energy computed with PIMC using two
         different nodal surfaces: the free particle density
         matrix and the variational density matrix.}
\label{energy_nodes}
\end{figure}
The differences in the internal energy, as shown in
Fig.~\ref{energy_nodes}, using FP and VDM nodes are significantly
smaller than the pressure deviations. One either finds agreement
within the error bars or that VDM nodes predict lower internal
energies, which was used in \cite{MC00} to show that VDM nodes are the
more accurate nodal surface. The observed energy differences did not
exceed $0.1\,\rm eV$ per atom.

For even lower density, the nodes are less relevant because they
become only important in a collision of two molecules, which occur
less frequently at lower density. This trend can also be understood in
terms of the degeneracy of the electrons. The degree of degeneracy
manifests itself in the path integral formalism by the probability for
the electrons to be involved in a permutation.  At high temperature,
the paths are very short and permutations are rare. At low temperature
and high density, the paths are long and can form long permutation
cycles. However in hydrogen at low density, the paths are localized
due to the attraction in atoms and molecules and permutations are
rare. Fig.~\ref{phase} shows that the permutation probability never
reaches $1\%$ for $r_s=4$ (see Tab.~\ref{table1}).  For higher
densities, the permutation probability is increased as indicated by
the contour lines. This is consistent with the temperature and density
dependence of the nodal error in pressure and internal energy
discussed above.

\section{Results}
\subsection{Equation of State}
\label{Equation_of_State}

\begin{figure}[htb]
\includegraphics[angle=0,width=\figurewidth]{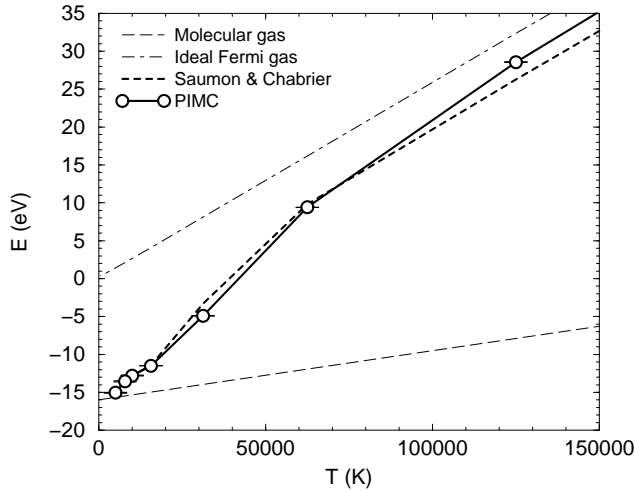}
\caption{Internal energy per atom vs. temperature for
         a density of $r_s=10$ comparing the SC-EOS \cite{SC92}
         with PIMC calculations.}
\label{E_vs_T_10.0_high}
\end{figure}

\begin{figure}[htb]
\includegraphics[angle=0,width=\figurewidth]{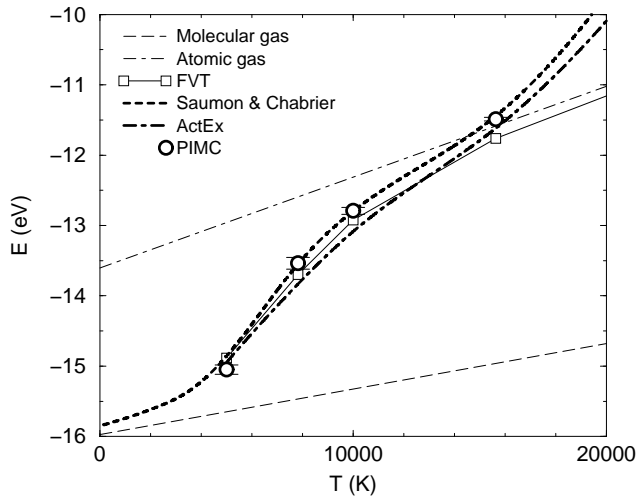}
\caption{Internal energy per atom vs. temperature for a
         density of $r_s=10$ as shown in Fig.~\ref{E_vs_T_10.0_high}
         but here for lower temperatures also including results from the
         activity expansion~\cite{Ro90}.}
\label{E_vs_T_10.0_low}
\end{figure}

\begin{figure}[htb]
\includegraphics[angle=0,width=\figurewidth]{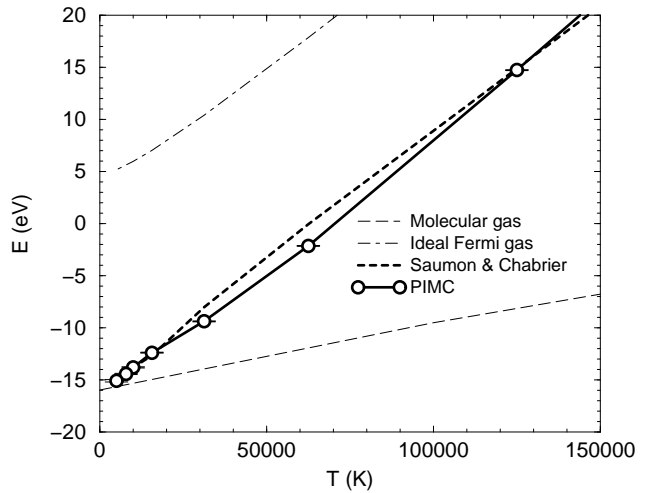}
\caption{Internal energy per atom vs. temperature as shown in
     Fig.~\ref{E_vs_T_10.0_high} but here for a density of $r_s=2.6$.}
\label{E_vs_T_2.6_high}
\end{figure}

\begin{figure}[htb]
\includegraphics[angle=0,width=\figurewidth]{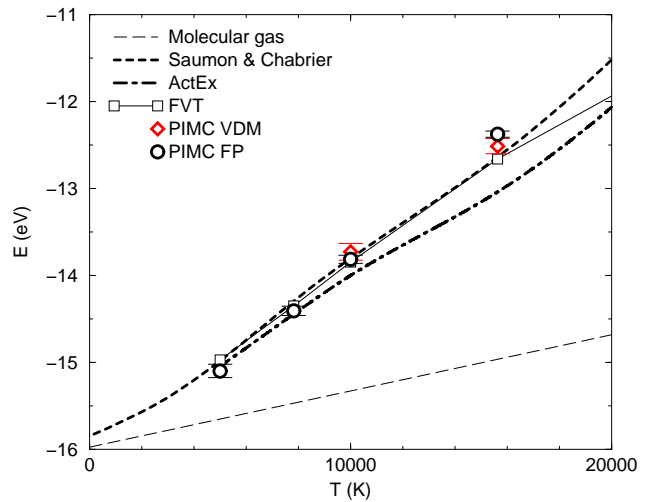}
\caption{Internal energy per atom vs. temperature for a
         density of $r_s=2.6$ as shown in Fig.~\ref{E_vs_T_2.6_high}
         but here for lower temperatures also including results from
         the activity expansion (ActEx)~\cite{Ro90} and the fluid
         variational theory (FVT)~\cite{Ju00}.}
\label{E_vs_T_2.6_low}
\end{figure}

Tab.~\ref{table1} gives the complete set of energies and pressures at
6 densities and 8 temperatures.  We now compare these results with
several models for hydrogen. We begin our discussion by studying the
internal energy per atom as a function of temperature shown in
Figs.~\ref{E_vs_T_10.0_high}~-~\ref{E_vs_T_2.6_low} for two selected
densities corresponding to $r_s=10.0$ and $2.6$. Generally, we find a
fairly good overall agreement with the EOS by SC
\cite{SC92} over the entire temperature and density range
discussed in this work. The agreement is particularly good in the
molecular and atomic regime for $r_s=10.0$, as shown in
Fig.~\ref{E_vs_T_10.0_low}. There the SC energies are within the
error bars of the PIMC results. At higher temperature shown in
Fig.~\ref{E_vs_T_10.0_high}, we find systematic deviations of up
to 5~eV per atom at $T=250\,000\,\rm K$. They indicate that the SC
energies are too low at high temperatures and too high at
intermediate temperatures (see Fig.~\ref{E_vs_T_10.0_high}). One
possible explanation for the deviations at high temperature is
that the SC model underestimates the degree of ionization (see
discussion in \cite{Sc95}).

We also studied these deviations as a function of density. The
cross-over temperature, above which the SC-EOS underestimates the
energy, increases with density. At $r_s=10.0$, the cross-over is
near $70\,000\,\rm K$ compared to $130\,000\,\rm K$
(Fig.~\ref{E_vs_T_2.6_high}) at $r_s=2.6$. At temperatures below
$20\,000\,\rm K$ for $r_s=2.6$, one also finds some small
deviations up to $0.5\,\rm eV$ per atom
(Fig.~\ref{E_vs_T_2.6_low}).

\begin{figure}[htb]
\includegraphics[angle=0,width=\figurewidth]{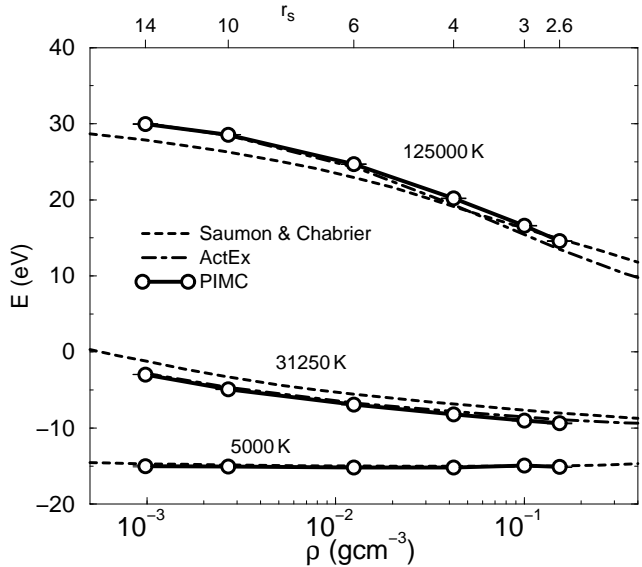}
\caption{Internal energy per atom vs. density for different temperatures
         from SC-EOS~\cite{SC92}, the activity expansion
         (ActEx)~\cite{Ro90} (not shown for $5000\,\rm{K}$ since
         nearly identical to SC) and PIMC calculations .}
\label{E_vs_rho}
\end{figure}

Fig.~\ref{E_vs_rho} shows a comparison of energy vs. density for
several temperatures. It shows that the SC-EOS overestimates the
energy for $T=5000\,\rm K$ and $31\,250\,\rm K$ and underestimates it
for $125\,000\,\rm K$ for densities higher than those corresponding to
$r_s=2.6$.

Now let us compare the pressure from the SC-EOS with that from the
PIMC simulation using Eq.~\ref{virial} in Tab.~\ref{table1}.
\begin{figure}[htb]
\includegraphics[angle=0,width=\figurewidth]{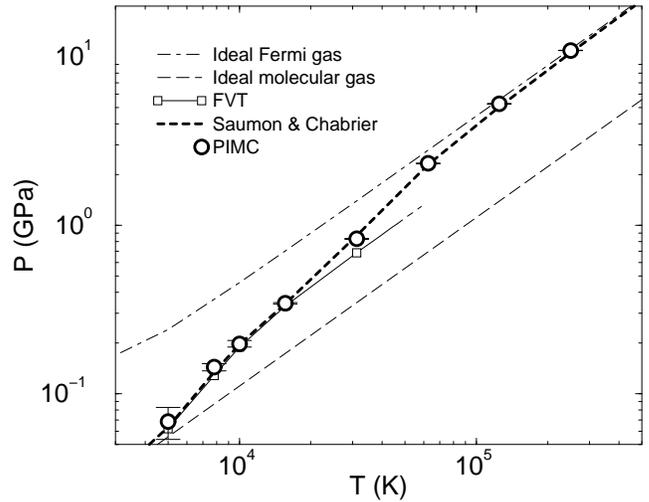}
\caption{Pressure vs. temperature at a density of $r_s=10$
         showing results from the fluid variational
         theory~\cite{Ju00}, the SC-EOS~\cite{SC92}, and PIMC
         simulations.}
\label{P_vs_T_10.0}
\end{figure}
We find remarkably good agreement of the entire range of
temperature and density under consideration. For a low density
such as $r_s=10.0$, this is shown in Fig.~\ref{P_vs_T_10.0}. As
expected, one finds that both methods interpolate between the
limit of an ideal Fermi gas at high temperatures and
non-interacting molecular gas at low temperatures.
Fig.~\ref{P_vs_T_2.6} confirms the good agreement at a higher
density of $r_s=2.6$. As a result of the strong interactions at
this density, one finds that the pressure at low temperatures is
significantly above the non-interacting molecular gas limit.

We find that the SC-EOS underestimates the pressure by about 3\%
for $T>62\,500\,\rm K$. This difference is outside the error bar
from the approximations in PIMC discussed in
section~\ref{Accuracy_of_the_Method} and could be interpreted as a
further indication, in addition to the observed energy deviations,
that the SC model underestimates the degree of ionization at high
temperatures. For intermediate temperatures $62\,500 \geq T \geq
15\,625\,\rm K$, one finds pressure differences, which are of the
same magnitude as the finite size effects in PIMC. For
temperatures below $15\,625\,\rm K$, the increased statistical
errors in the PIMC pressure are of the same size as the observed
deviations.

\begin{figure}[htb]
\includegraphics[angle=0,width=\figurewidth]{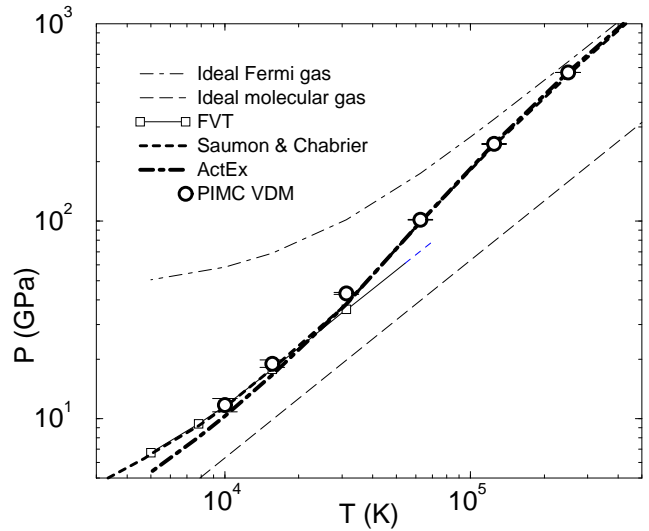}
\caption{Pressure vs. temperature at a density of $r_s=2.6$
         as shown in Fig.~\ref{P_vs_T_10.0}.}
\label{P_vs_T_2.6}
\end{figure}

\begin{figure}[htb]
\includegraphics[angle=0,width=\figurewidth]{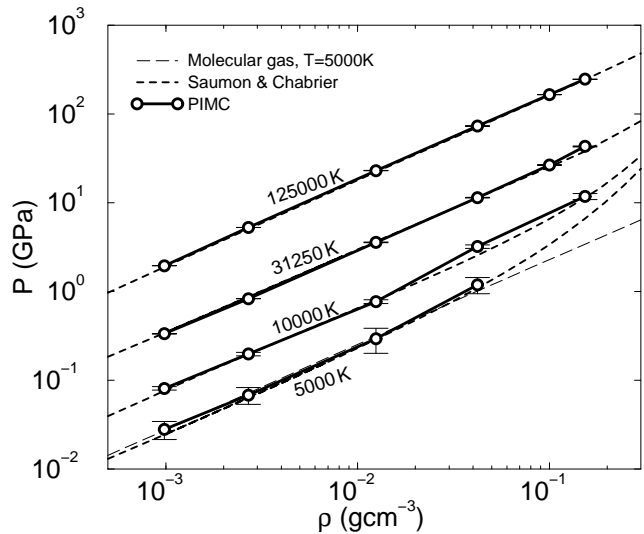}
\caption{Pressure vs. density for different temperatures.}
\label{P_vs_rho}
\end{figure}

In Fig.~\ref{P_vs_rho}. we show the pressure as a function of density,
which confirms the good agreement. The figure also indicates that, at
$5000\,\rm K$ and $r_s \geq 4$, the pressure is close to the pressure
of a non-interacting molecular gas.

\begin{table}
\caption{Pressure and internal energy per atom and the resulting Hugoniot from PIMC
         simulations with $32$ pairs of particles and $\tau^{-1}=2
         \cdot 10^6\,\rm{K}$ using free particle nodes except for $^*$
         where VDM nodes were employed instead. The probabilities $x$
         for finding a proton in a given state for the three dominant
         species are derived from a cluster analysis. $P_{\rm perm}$
         is the permutation probability for the electrons.}
\begin{tabular}{rrrrrrrrr}
$r_s$ & $T ({\rm K})$ & $P\,(\rm{GPa})~~~$ & $E\,(\rm{eV})~~~$ & $x_{\rm H^+}$ & $x_{\rm H}$ & $x_{\rm H_2}$ & $P_{\rm perm}$\\
\colrule
14.0 & 250$\,$000&    4.000$\,$(2)$\;\;$  &  62.93$\,$(4)$\;\;\;\;$  &       &      &          & 0.000\\
14.0 & 125$\,$000&    1.955$\,$(2)$\;\;$  &  29.97$\,$(4)$\;\;\;\;$  & 0.99 & 0.01  &   0.00   & 0.000\\
14.0 &  62$\,$500&    0.901$\,$(2)$\;\;$  &  11.85$\,$(4)$\;\;\;\;$  & 0.95 & 0.05  &   0.00   & 0.000\\
14.0 &  31$\,$250&    0.334$\,$(2)$\;\;$  &  --2.97$\,$(3)$\;\;\;\;$ & 0.64 & 0.36  &   0.00   & 0.000\\
14.0 &  15$\,$625&    0.127$\,$(2)$\;\;$  & --11.30$\,$(4)$\;\;\;\;$ & 0.24 & 0.76  &   0.00   & 0.000\\
14.0 &  10$\,$000&    0.081$\,$(4)$\;\;$  & --12.43$\,$(6)$\;\;\;\;$ & 0.12 & 0.72  &   0.15   & 0.000\\
14.0 &   7$\,$812&    0.047$\,$(5)$\;\;$  & --13.34$\,$(13)$\;\;$    & 0.08 & 0.58  &   0.33   & 0.000\\
14.0 &   5$\,$000&    0.028$\,$(6)$\;\;$  & --15.00$\,$(12)$\;\;$    & 0.01 & 0.11  &   0.88   & 0.000\\
\colrule
10.0 & 250$\,$000&  10.902$\,$(4)$\;\;$   &  62.00$\,$(3)$\;\;\;\;$  &      &       &          & 0.000\\
10.0 & 125$\,$000&   5.259$\,$(6)$\;\;$   &  28.56$\,$(3)$\;\;\;\;$  & 0.98 & 0.02  &   0.00   & 0.000\\
10.0 &  62$\,$500&   2.329$\,$(5)$\;\;$   &   9.41$\,$(3)$\;\;\;\;$  & 0.90 & 0.10  &   0.00   & 0.000\\
10.0 &  31$\,$250&   0.831$\,$(5)$\;\;$   &  --4.91$\,$(3)$\;\;\;\;$ & 0.57 & 0.43  &   0.00   & 0.000\\
10.0 &  15$\,$625&   0.344$\,$(4)$\;\;$   & --11.49$\,$(3)$\;\;\;\;$ & 0.23 & 0.74  &   0.02   & 0.000\\
10.0 &  10$\,$000&   0.198$\,$(9)$\;\;$   & --12.79$\,$(5)$\;\;\;\;$ & 0.11 & 0.60  &   0.27   & 0.000\\
10.0 &   7$\,$812&   0.144$\,$(7)$\;\;$   & --13.54$\,$(9)$\;\;\;\;$ & 0.03 & 0.34  &   0.62   & 0.000\\
10.0 &   5$\,$000&   0.068$\,$(15)   & --15.05$\,$(7)$\;\;\;\;$      & 0.00 & 0.07  &   0.92   & 0.000\\
\colrule
6.0 & 250$\,$000& 49.46$\,$(3)$\;\;$ &  59.33$\,$(4)$\;\;\;\;$   &      &       &         & 0.000\\
6.0 & 125$\,$000& 23.00$\,$(3)$\;\;$ &  24.70$\,$(4)$\;\;\;\;$   & 0.94 & 0.06  &   0.00  & 0.000\\
6.0 &  62$\,$500&  9.56$\,$(2)$\;\;$ &   4.79$\,$(3)$\;\;\;\;$   & 0.80 & 0.20  &   0.00  & 0.000\\
6.0 &  31$\,$250&  3.58$\,$(3)$\;\;$ &  --6.92$\,$(4)$\;\;\;\;$  & 0.52 & 0.46  &   0.01  & 0.000\\
6.0 &  15$\,$625&  1.52$\,$(2)$\;\;$ & --11.83$\,$(4)$\;\;\;\;$  & 0.21 & 0.68  &   0.09  & 0.001\\
6.0 &  10$\,$000&  0.77$\,$(4)$\;\;$ & --13.38$\,$(6)$\;\;\;\;$  & 0.08 & 0.49  &   0.40  & 0.000\\
6.0 &   7$\,$812&  0.63$\,$(5)$\;\;$ & --13.95$\,$(7)$\;\;\;\;$  & 0.04 & 0.38  &   0.56  & 0.000\\
6.0 &   5$\,$000&  0.29$\,$(9)$\;\;$ & --15.17$\,$(12)$\;\;$     & 0.00 & 0.09  &   0.90  & 0.000\\
\colrule
4.0 & 250$\,$000& 162.46$\,$(10)$\;\;$ &      55.63$\,$(4)$\;\;\;\;$  &      &       &         & 0.000 \\
4.0 & 125$\,$000&  73.00$\,$(23)$^*$ &  20.24$\,$(9)$^*\;\;$          & 0.88 & 0.12  &   0.00  & 0.000\\
4.0 &  62$\,$500&  29.75$\,$(16)$^*$ &   1.23$\,$(6)$^*\;\;$          & 0.72 & 0.26  &   0.00  & 0.001\\
4.0 &  31$\,$250&  11.22$\,$(22)$^*$ &  --8.32$\,$(8)$^*\;\;$         & 0.47 & 0.46  &   0.03  & 0.004\\
4.0 &  15$\,$625&   5.01$\,$(17)$^*$ & --11.87$\,$(6)$^*\;\;$         & 0.19 & 0.58  &   0.18  & 0.008\\
4.0 &  10$\,$000&   3.23$\,$(30)$^*$ & --13.43$\,$(11)$^*$            & 0.03 & 0.30  &   0.63  & 0.005\\
4.0 &   7$\,$812&   2.20$\,$(14)$\;\;$ &     --14.29$\,$(6)$\;\;\;\;$ & 0.01 & 0.18  &   0.80  & 0.004\\
4.0 &   5$\,$000&   1.19$\,$(25)$\;\;$ &     --15.20$\,$(9)$\;\;\;\;$ & 0.00 & 0.11  &   0.88  & 0.002\\
\colrule
3.0 & 250$\,$000& 374.47$\,$(14)$\;\;$ &  51.79$\,$(2)$\;\;\;\;$    &      &      &         & 0.000 \\
3.0 & 125$\,$000& 165.21$\,$(22)$\;\;$ &  16.58$\,$(4)$\;\;\;\;$    & 0.83 & 0.15 &    0.01 & 0.001\\
3.0 &  62$\,$500&  67.70$\,$(25)$\;\;$ &  --1.04$\,$(4)$\;\;\;\;$   & 0.66 & 0.29 &    0.01 & 0.005\\
3.0 &  31$\,$250&  26.67$\,$(24)$\;\;$ &  --9.02$\,$(4)$\;\;\;\;$   & 0.45 & 0.43 &    0.06 & 0.028\\
3.0 &  15$\,$625&  13.08$\,$(28)$\;\;$ & --12.29$\,$(4)$\;\;\;\;$   & 0.15 & 0.42 &    0.34 & 0.059\\
3.0 &  10$\,$000&   9.26$\,$(26)$\;\;$ & --13.79$\,$(4)$\;\;\;\;$   & 0.03 & 0.35 &    0.60 & 0.037\\
\colrule
2.6 & 250$\,$000& 566.4$\,$(4)$\;\;$     &  49.58$\,$(4)$\;\;\;\;$  &      &      &        & 0.000 \\
2.6 & 125$\,$000& 246.0$\,$(5)$^*$&  14.57$\,$(5)$^*\;\;$           & 0.80 & 0.17 &   0.01 & 0.002\\
2.6 &  62$\,$500& 101.7$\,$(4)$^*$&  --2.25$\,$(4)$^*\;\;$          & 0.65 & 0.28 &   0.02 & 0.014\\
2.6 &  31$\,$250&  43.1$\,$(5)$^*$&  --9.38$\,$(5)$^*\;\;$          & 0.42 & 0.38 &   0.09 & 0.078\\
2.6 &  15$\,$625&  19.0$\,$(8)$^*$& --12.51$\,$(9)$^*\;\;$          & 0.14 & 0.42 &   0.33 & 0.168\\
2.6 &  10$\,$000&  11.7$\,$(9)$^*$& --13.73$\,$(10)$^*$             & 0.02 & 0.34 &   0.60 & 0.126\\
\end{tabular}
\label{table1}
\end{table}
In our comparison, we also included results from the activity
expansion by Rogers~\cite{Ro90}, which shows very good agreement
in pressure and internal energy (see Figs.~\ref{E_vs_T_10.0_low},
\ref{E_vs_T_2.6_low}, and \ref{E_vs_rho}).  The differences are
small but increase with density.  In the molecular and atomic
regime, one also finds good agreement with the fluid variational
theory (FVT) by Juranek and Redmer \cite{Ju00} as shown in
Figs.~\ref{E_vs_T_10.0_low}, \ref{E_vs_T_2.6_low},
\ref{P_vs_T_10.0}, and \ref{P_vs_T_2.6}. For higher temperatures,
the FVT model is not applicable since it does not include
ionization of atoms.

\subsection{Pair Correlation Functions}
\label{Pair_Correlation_Functions}

\begin{figure}[htb]
\includegraphics[angle=0,width=\figurewidth]{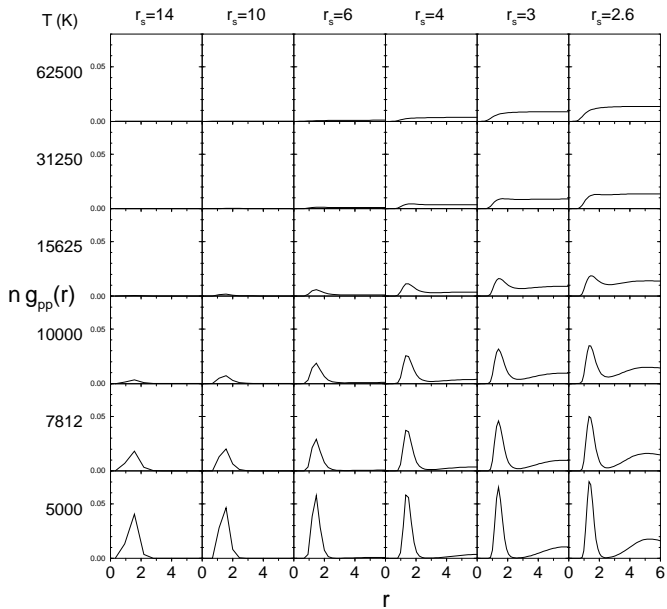}
\caption{Proton-proton pair correlation function multiplied by the density $n$.
         The columns correspond to to $r_s$ values and the rows to different temperatures $T$.}
\label{grpp}
\end{figure}

There are four different pair correlation
functions which can be directly obtained from many-body
simulations and provide direct information about the state of the
system. Shown in the following figures is an extensive set of
pair correlations which allow one to estimate the microscopic
structure of the system and allow a direct comparison with other
simulations. The proton-proton pair correlation functions from
PIMC simulations with free particle nodes are shown in
Fig.~\ref{grpp}. For $T \leqs 20\,000\,\rm K$ a peak at the bond
length of $1.4$ emerges, which clearly demonstrates the formation
of molecules. We found it useful to multiply the pair correlation
function by an extra density factor $n=N/v$ so that the area under
the peak is proportional to the molecular fraction. The peak
height gets smaller with decreasing density as a result of
entropic dissociation of the molecules, driven by the number of
unbound states at low density.  Thermal dissociation also reduces
the number of molecules with increasing temperature. For
$r_s\leqs2$, we expect that pressure dissociation diminishes the
number of molecules with increasing density \cite{BM00} but this
density range is beyond the scope of this paper.

\begin{figure}[htb]
\includegraphics[angle=0,width=\figurewidth]{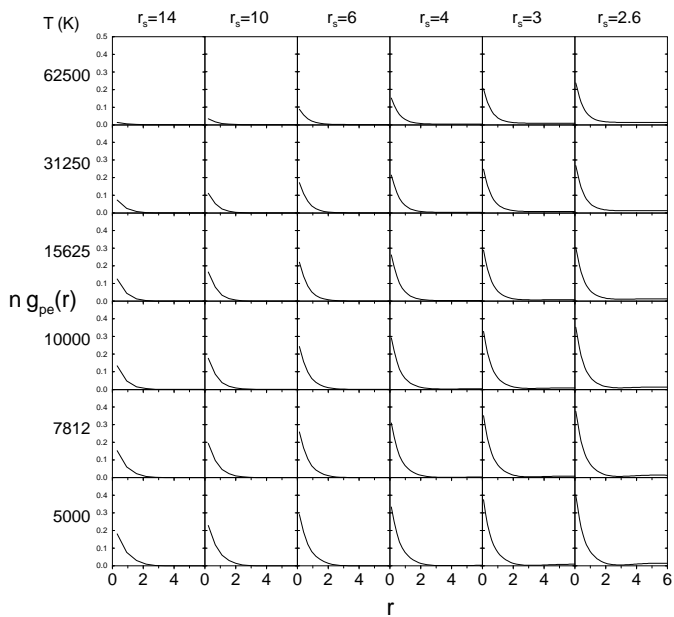}
\caption{Proton-electron pair correlation function multiplied by the density $n$.
         The columns correspond to different $r_s$ values and the rows to different temperatures $T$.}
\label{grpe}
\end{figure}

The proton-electron pair correlation function multiplied by the
density is shown in Fig.~\ref{grpe}. The peak near the origin
shows the increased probability of finding an electron near a
proton due to the Coulomb attraction. The peak height decreases
with temperature and increases with density because of thermal
ionization and entropy ionization respectively. At low
temperature, the peak can be interpreted as occupation of bound
states although (unbound) scattering states can also contribute.
From proton-electron pair correlation alone, one cannot
distinguish between an atomic and a molecular state.

\begin{figure}[htb]
\includegraphics[angle=0,width=\figurewidth]{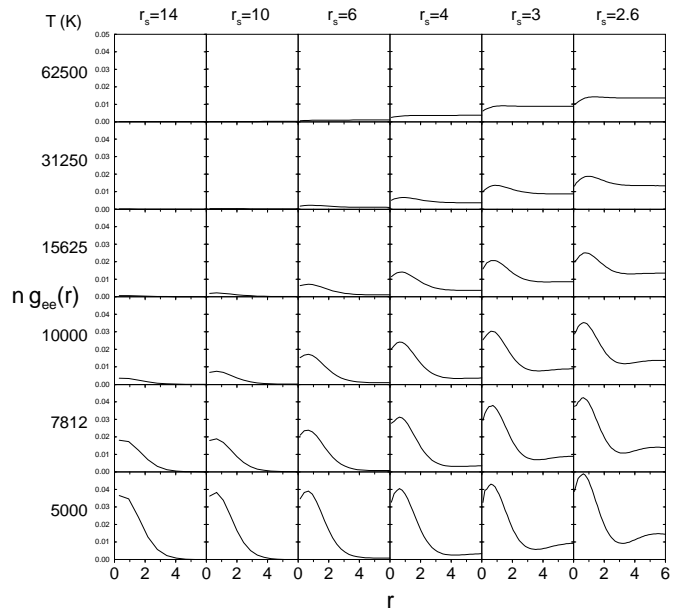}
\caption{Electron-electron pair correlation function for electrons
with opposite
         spin multiplied by the density $n$ is shown. The columns
         correspond to different $r_s$ values and the rows to different
         temperatures $T$.}
\label{gree}
\end{figure}

\begin{figure}[htb]
\includegraphics[angle=0,width=\figurewidth]{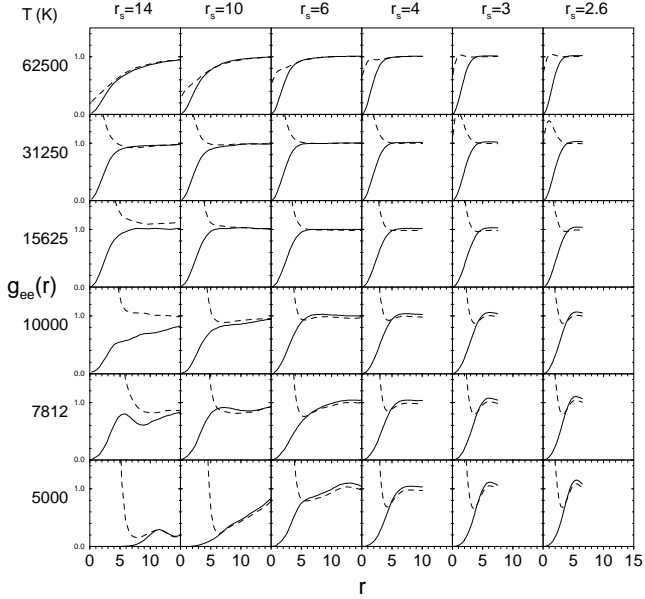}
\caption{Electron-electron pair correlation function. The solid lines
         correspond to pairs of electrons with parallel spin. For the
         sake of comparison, we also show the pair correlation
         functions of pairs with opposite spin as dashed lines. This
         function is strongly peaked near the origin in the presence
         of molecules as shown in Fig.~\ref{gree}.  The columns
         correspond to different $r_s$ values and the rows to
         different temperatures $T$.}
\label{gree2}
\end{figure}

Figs.~\ref{gree} and \ref{gree2} show the electron-electron pair
correlation functions for pairs with anti-parallel spin. The peak
at small separations comes from the formation of the molecular
bond. For pairs of electrons with parallel spin, one always finds
a strong repulsion due to the Pauli exclusion principle and to a
lesser extent to the Coulomb repulsion. This is shown in
Fig.~\ref{gree2}.

\subsection{State of Hydrogen}

In this section, we discuss the {\it phase diagram} of hydrogen as
shown in Fig.~\ref{phase}.  The diagram shows the approximate
location of the molecular, the atomic and the plasma regimes. The
PIMC simulations, since they are based on the basic description in
terms of electrons and protons, do not directly lend themselves to
determining the number of compound particles such as molecules and
atoms. (Methods for determining this from PIMC simulations will be
discussed in a future publication.) In order to obtain an estimate
of the atomic and molecular fraction, we employed a cluster
analysis of the PIMC path configurations. As described in
\cite{Mi98}, we consider two protons as belonging to one cluster
if they are less than $1.9\,a_{\rm B}$ apart. An electron belongs to one
particular cluster if it is less than $1.4\,a_{\rm B}$ away from any proton
in the cluster. The two cut-off radii were chosen from the molecular
and atomic ground state distribution. This analysis gives reasonable
estimates for the molecular and atomic fractions at low
temperatures. At high temperature, it overestimates the number of
bound states because even in an (unbound) collision, two particles are
counted falsely as being part of the cluster. We corrected for this by
applying an additional criterion: a particle can only be considered as
bound if the difference in action to remove it from the cluster is
positive. This method leads to the expected corrections at high
temperature. (The regime boundaries in Fig.~\ref{phase} discussed
below are hardly affected by the additional correction.)

The lower dashed line represents the region where 60\% of the
protons are bound in molecules. When the number of protons bound
in atoms ( {\it i.e.}  with an electron) drop below 40\% we
labeled this state as plasma as shown in the upper dashed-dotted
line. It should be emphasized that the location of these lines
depends on the choice of these limits as well as on the cut-off
radii used to determine the clusters in this place.
Fig.~\ref{phase} also shows the location of isobars, which appear
as almost straight lines in this double logarithmic graph. The
slope is different from the ideal gas because the pressure depends
on ionization and dissociation.

Tab.~\ref{table1} shows the fraction of the three most frequently
found species: molecules, atoms and free protons where $x$ is
defined as the probability of finding a proton in a certain
compound particle. It should be noted that the sum: $x_{\rm
H^+}$+$x_{\rm H}$+$x_{\rm H_2}$ is less than 1 since other
clusters have a non-zero probability. The largest contributions
besides those listed are H$_2^+$ with a maximum of 0.06 for
$r_s=2.6$ and $T=15\,625\,\rm K$ followed by H$_3$ with $x \leq
0.03$ and H$^-$ with $x \leq 0.02$. Even larger clusters occur
very infrequently. The cluster analysis also gives an estimate for
the fraction of free electrons, which agrees well with the number
of ionized protons: $x_{\rm H^+}$.

\begin{figure}[htb]
\includegraphics[angle=0,width=\figurewidth]{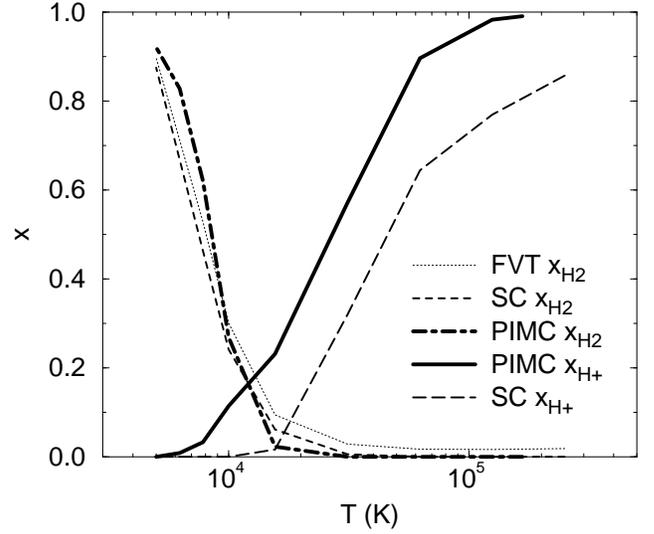}
\caption{Fraction of molecules and free protons as a function of
    temperature for $r_s=10$ comparing the cluster analysis of the
    PIMC results with the SC model~\cite{SC92} and the fluid
    variational theory (FVT)~\cite{Ju00}.}
\label{cluster_10.0}
\end{figure}

\begin{figure}[htb]
\includegraphics[angle=0,width=\figurewidth]{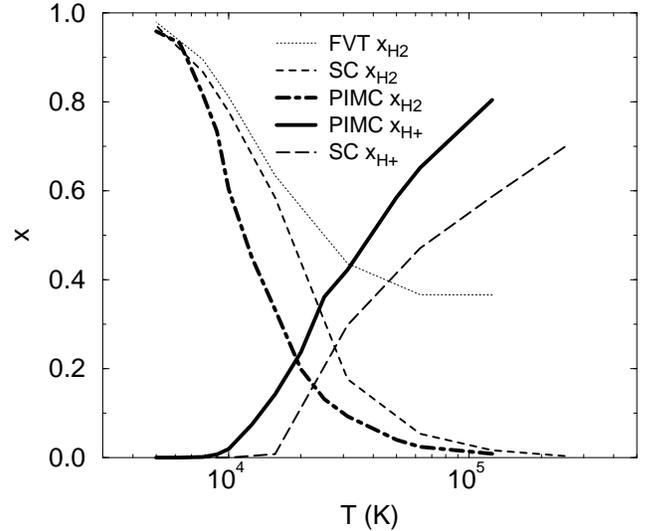}
\caption{Fraction of molecules and free protons as shown in
         Fig.~\ref{cluster_10.0} but here for a density of $r_s=2.6$.}
\label{cluster_2.6}
\end{figure}

Fig.~\ref{cluster_10.0} shows a comparison of the fraction of
molecules and ionized atoms for $r_s=10$. One finds that the molecular
fractions decays rapidly with temperature. The resulting atoms are
then ionized at even higher temperatures leading to the observed
increase in the number of free protons. The PIMC predictions for the
molecular fraction agree very well with the SC model as well as with
the FVT. On the other hand, the PIMC results shows a significantly
higher degree of ionization than SC.
The same comparison for a higher density of $r_s=2.6$ in
Fig.~\ref{cluster_2.6} shows that the cluster analysis leads to smaller
number of molecules than predicted by the SC model.

Summarizing one can say that the cluster analysis provides us with
reasonable estimates for the number of atoms and molecules in the
considered density range. We caution the reader that other definitions
of atoms and molecules, possible in a many-body hydrogen, while giving
qualitatively similar results, may differ quantitatively. Our computed
numerical values must be used with caution. A rigorous, many-body
definition of a bound state remains to be applied. Several ideas are
discussed in \cite{Gi90}.

\section{Conclusions}
In this work, we studied the high-temperature equation of state of
hydrogen at low and intermediate densities and find a remarkably good
agreement with the SC-EOS. Generally, one finds that the deviations in
the energy are more pronounced than the differences in the
pressure. We find significant deviations in the EOS of temperatures
$\approx 100 000 K$, most likely caused by an underestimate of the
degree of ionization at those temperatures.  In future work, we will
extend this comparison to higher densities. There one expects to find
substantial differences between the SC model and PIMC, which manifest
themselves in a different shock Hugoniot~\cite{MC00}.

\begin{acknowledgments}
The authors would like to thank E.L.~Pollock for useful discussions as
well as D.~Saumon, F.~Rogers, and H.~Juranek for providing us with
their EOS tables.
This research was funded by the U.S. Department of Energy through the
University of California under Subcontract number B341494 and in part
by the University of California/Lawrence Livermore National Laboratory
(LLNL) under contract no. W-7405-Eng-48.
We used the computational facilities at the LLNL and at the National
Center for Supercomputing Applications.
\end{acknowledgments}
%
%
\widetext

\end{document}